\documentclass[12pt,aps,pra,english,prstper,reprint,raggedbottom, superscriptaddress,tightenlines,longbibliography,twocolumn]{revtex4-1}   
\usepackage{graphicx,color,braket,footnote,subfigure}
\usepackage{amsmath,amsthm,bm,bbm,bmpsize-base}
\usepackage[T1]{fontenc}
\usepackage[utf8]{inputenc}
\usepackage{amssymb}
\usepackage{natbib}
\newcommand{\bla}{\color{black}}

\usepackage{dsfont,hyperref,xcolor}
\hypersetup{colorlinks,linkcolor={red},citecolor={blue},urlcolor={red}}
\usepackage{amsxtra}
\DeclareMathOperator{\sech}{sech}

\begin{document}
	\title{On the non-Markovianity of quantum semi-Markov processes }
	\author{
		Shrikant Utagi}  
	\affiliation{Department of Theoretical Sciences, Poornaprajna    Institute   of   Scientific
		Research, Bengaluru -  562164, India} 
	\affiliation{Optics and Quantum Information Group, The Institute of Mathematical Sciences, CIT Campus, Taramani, Channai, India.}
	\author{Subhashish  Banerjee}
	\email{subhashish@iitj.ac.in}
	\affiliation{Indian Institute of Technology Jodhpur,  Jodhpur- 342011, India}
	
	\author{R.     Srikanth}  
	\email{srik@poornaprajna.org}
	\affiliation{Department of Theoretical Sciences, Poornaprajna    Institute   of
		Scientific     Research,     Bengaluru      -     562164,     India}

	\begin{abstract}
			The non-Markovianity of the stochastic process called the quantum semi-Markov (QSM) process is studied using a recently proposed quantification of memory based on the deviation from semigroup evolution, that provides a unified description of divisible and indivisible channels. This  is shown to bring out the property of QSM processes to exhibit memory effects even in the CP-divisible regime, in agreement with an earlier result. An operational meaning to the non-Markovian nature of semi-Markov processes is also provided.
%
%
	\end{abstract}
	\maketitle
	
	\section{Introduction}
	Open quantum systems is the study of the evolution of the system of interest taking into account the  effect of its environment \cite{banerjee2018open,breuer2002theory}. The effect of the environment is encoded in the nature of its interaction with the system of interest \cite{grabert1988quantum,sbqbm,banerjee2000quantum}.  Traditionally, weak system environment interaction is associated with Markovian (memoryless) regime, where the environmental time scale is much smaller than the system time scale.  A more involved scenario occurs when the effects of memory are considered, and is broadly called non-Markovian dynamics. Defining, characterizing, detecting and quantifying non-Markovianity of open system dynamics has been an intense research activity in the last decade \cite{RHP14,breuer2016colloquium,vega2017dynamics,li2018concepts,shrikant2020temporal,shrikant2018non-Markovian}. This has impacted a wide range of applications, ranging from quantum cryptography \cite{vasile2011continuous,thapliyal2017quantum,shrikant2020pingpong}, quantum walks \cite{kumar2017nonmarkovian,naikoo2020non}, quantum thermodynamics \cite{thomas2018thermodynamics}, quantum coherence and correlations \cite{laine2014nonlocal,bhattacharya2018evolution,banerjee2010dynamics,paulson2020hierarchy}. 
	
	Historically speaking, a process' deviation from being a semigroup was associated with non-Markovianity \cite{breuer2016colloquium}. Later, in the quantum dynamical maps approach, one discarded this notion in favour of a stronger definition of completely positive (CP-) divisibility \cite{RHP10}.  A revisit to the historical definition was proposed in  \cite{shrikant2020temporal}, where a quantitative notion of non-Markovianity was developed by identifying it with deviation from ``temporal self-similarity'', the property of a system dynamics whereby the propagator between two intermediate states is independent of the initial time, which was then identified with the quantum dynamical semigroup (QDS). 
In each of the number of approaches devoted to the study of open system dynamics, such as the GKSL master equation approach \cite{lindblad1976generators,gorini1976completely}, collisional models \cite{vacchini2014general,ciccarello2013collision}, and process tensor formalism \cite{pollock2018operational}, the way (non-)Markovianity is defined tends to vary.  Another approach was made in \cite{shabani2005completely} by introducing a post-Markovian master equation. 
These models aren't equivalent: for example, a process that is non-Markovian according to the CP-divisibility condition \cite{RHP10} may not be non-Markovian according to the distinguishability or information back-flow criterion \cite{breuer2009measure}. 
	
As mentioned earlier, a universal definition of non-Markovianity has remained elusive and in this light there have been recent proposals of reinstating a definition that accounts for memory effects present even in CP-divisible processes. See for example, a definition of non-Markovianity introduced based on conditional past-future (CPF) correlations \cite{budini2018quantum,budini2019conditional} and also the one based on deviation from temporal self-similarity of map \cite{shrikant2020temporal}. Also, one may define a hierarchy of divisibility of quantum dynamical maps, namely P-divisibility \cite{chruscinski2014degree}, CP-divisibility \cite{RHP10} and L-divisibility \cite{davalos2019divisibility}. The latter has been shown to be equivalent to the semigroup evolution. In this light, one may argue that L-divisibility, lack of CPF correlations, and temporal self-similarity of dynamical maps, are equivalent definitions pertaining to the semigroup evolution, while deviation from any of these notions signals non-Markovian evolution of the weakest form, even when the dynamics is CP-divisible.

A class of quantum processes are the so-called semi-Markov processes \cite{breuer2009structure} whose dynamical maps can deviate from being a one-parameter semigroup. In classical statistics, the term ``semi-Markov'' denotes continuous-time processes characterized by jump probabilities between sites (that a system can occupy) and waiting time distributions (WTDs) determining the time between jumps. The process reduces to a Markov chain when the WTD is an exponential distribution, and is non-Markovian for general WTDs, since the prior time that has been spent in the state can manifest in the subsequent statistics as a memory effect. The quantum semi-Markov process is the quantum analogue of the classical case, with the jumps realized by CPTP maps acting on the reduced state of the system \cite{megier2021memory}. Noteworthy in the present context is that quantum semi-Markov processes can be CP-divisible in certain parametric regimes and CP-indivisible in others. In this work, we show that the memory in the semi-Markov process can also be indicated and quantified using the temporal self-similarity criterion of Ref. \cite{shrikant2020temporal}. This provides a new perspective on the memory effect of semi-Markov processes. This approach was earlier applied to a class of noises called the Power Law (PL) and Ornstein-Uhlenbeck (OU) noises that have memory kernel with colored spectral density but are Markovian according to the CP-divisibility condition \cite{kumar2017nonmarkovian}. Thus, for example, this criterion can be used to indicate the memory of OU noise, which is \textit{operationally} non-Markovian in the sense that it can delay entanglement sudden death in an entangled two-spin system evolution \cite{yu2010entanglement}. 
	
	The plan of the paper is as follows. In Section \ref{sec:sssmeasure} we review a measure of non-Markovianity introduced in \cite{shrikant2020temporal}. In Section \ref{sec:sec2}, we briefly review classical and quantum semi-Markov processes, particularly focusing on dephasing quantum semi-Markov processes.  In Section \ref{sec:sec4}, we argue that quantum semi-Markovian processes that are CP-divisible can be non-Markovian and quantify the non-Markovianity of these processes using a recently introduced measure \cite{shrikant2020temporal}. In Section \ref{sec:sec3}, we point out the possible usefulness of quantum semi-Markov processes from an information theoretic perspective. Then we conclude in Section \ref{sec:sec5}.

\section{Measure of non-Markovianity based on the deviation from temporal self-similarity of the map \label{sec:sssmeasure}}
Here, we briefly review the definition and quantification of quantum memory effects based on the temporal self-similarity criterion \cite{shrikant2020temporal}. This quantification can be used to indicate a weak form of memory even in case of CP-divisible dynamics, and will be applied to semi-Markov processes in a later section. 

A generic two-parameter dynamical map is given by $\Phi(t, t_i) = \mathcal{T} \text{exp}\{ \int_{t_i}^{t} \mathcal{L}(\tau) d\tau \}$ with $t_i < \tau < t$ where $\mathcal{T}$ is the time-ordering operator. Now, an \textit{infinitesimal} map may be given by $(\delta\Phi)\rho(t) = \mathcal{T}\exp\left(\int_t^{t+dt}\mathcal{L}(\tau)d\tau\right)\rho(t)=  (1+\mathcal{L}(t)dt)\rho(t)$.   Consider the quantity $\Vert \delta\Phi(t)- \delta\Phi \Vert = \Vert \mathcal{L}(t) dt - \mathcal{L} dt \Vert$, where $\Phi = \text{exp}\{t \mathcal{L}\}$ is the dynamical semigroup generated by a time-\textit{in}dependent generator $\mathcal{L}$. Here, $\Vert A \Vert = \sqrt{A A^\dagger}$ is the trace norm. Based on this, Ref. \cite{shrikant2020temporal} defines a measure of non-Markovianity as a function of the magnitute of deviation from semigroup evolution, which for brevity may be called the SSS measure (after the authors' names):
\begin{equation}
 \xi = \min_{\mathcal{L}} \frac{1}{T} \int_0^T \Vert \chi_{\mathcal{L}(t)} -  \chi_{\mathcal{L}}\Vert dt,
	\label{eq:fqds}
\end{equation}
where $\chi_{\bullet}  \equiv (\bullet \otimes \mathbb{I})\ket{\Psi}\bra{\Psi}$ and $\ket{\Psi}$ \bla is the unnormalized maximally entangled state $\ket{\Psi} \equiv \sum_j \ket{j,j}$. 
As a simple illustration, consider the measure applied to dephasing noise of a $d$-dimensional quantum system (qudit). Interestingly, the measure turns out to be independent of the system dimension in the context of Pauli channels. 

Consider a qudit depolarizing channel, described by $\mathcal{L}[\rho] = \sum_\alpha \frac{\gamma_\alpha(t)}{d}(\mathbb{U}_\alpha[\rho] -(d-1) \rho)$, where $\mathbb{U}_\alpha[\rho] = \sum_j A^j_\alpha \rho A^{j\dagger}_\alpha$, where $A_\alpha$ are the Pauli operators in prime power dimension $d$, and $\alpha \in \{1,2,3\}$. As a simple instance of applying the SSS measure, consider a family of dephasing channels $\mathcal{L}[\rho] = \frac{\gamma_z(t)}{d}(\mathbb{U}_z[\rho] - \rho)$. We have
\begin{align}
\xi &=  \min_{\mathcal{\mathcal{L}}}\frac{1}{T}\int_0^T \Vert \chi_{\mathcal{L}(t)} -  \chi_{\mathcal{L}}  \Vert dt \bla \nonumber \\ &=  \min_{\gamma}\frac{1}{Td}\int_0^T dt(\gamma_z(t) - \gamma) \nonumber\\
 &\times \Vert\sum_k (Z_d^k \otimes \mathds{1}_d) \ket{j,j}\bra{j,j} (Z_d^{k,\dagger} \otimes \mathds{1}_d) \Vert
 \nonumber\\
 &= \min_{\gamma}\frac{1}{T}\int_0^T dt \vert \gamma_z(t) - \gamma \vert,
\label{eq:ddimdephase}
\end{align}
where $\gamma_z(t)$ is the decay rate of the process in the region beyond the semigroup and $\gamma$ is the (time-independent) Lindblad rate characterizing the semigroup evolution. 


It is important to note here that this quantification of non-Markovianity is applicable quite generally, even to maps that are non-invertible, whereas one based on regions where a decoherence rate is negative \cite{hall2014canonical}, can overestimate the quantity, because positivity of the decoherence rates is only sufficient but not necessary in case of non-invertible maps \cite{chrusinski2018divisibility}.

	\section{Classical and quantum semi-Markov processes \label{sec:sec2}}
	A classical stochastic process is one which takes a valid probability distribution to another valid probability distribution. For example, consider a classical stochastic process \cite{vacchini2011markovianity} in which a system, characterized by a set of states, jumps from one state to another with a jump probability $\pi$, such that the dynamics of the system is given by 
	\begin{equation}
		Q(\tau) = \begin{pmatrix}
			1- \pi  & \pi \\
			\pi   &  1- \pi 
		\end{pmatrix} f(\tau) \equiv \Pi f(\tau),
	\end{equation}
	where $\Pi$ is the transition matrix and $f(\tau)$ is called the waiting time distribution. If the probability distribution which characterizes the time distribution of the system to stay in a particular state is given by an exponential probability distribution say $f(\tau) = \lambda e^{-\lambda \tau}$, then the process is classically Markovian, and non-Markovian otherwise.  The survival probability is $g(t) = 1 - \int_{0}^{t} d\tau f(\tau)$.
	
	This classical concept, when transported to the quantum context in its weakest sense, gives the quantum semi-Markov (QSM) process, whose time-local and time-nonlocal master equations are studied in Ref. \cite{megier2020evolution}. 

Suppose a channel defines the CPTP projector $\mathcal{P}$, which is a CPTP map with the idempotency property, i.e., $\mathcal{P}\rho = \mathcal{P}^2\rho$. The generator of the corresponding open quantum system dynamics may be written as 
\begin{align}
		\mathcal{L}(t) = \gamma(t)[\mathcal{P}-\mathds{1}],
		\label{eq:gen}
	\end{align}
where, $\gamma(t)$ is the time-dependent decay rate. Equivalently, noting that the map $\Phi(t) = \mathcal{T} \exp \{\int_{t_0}^{t_{\rm max}}\mathcal{L}(\tau)d\tau\}$ itself obeys a non-local master equation \begin{align}
	\frac{d\Phi(t)}{dt} = \int_{0}^{t} \mathcal{K}(t-\tau)  \Phi(\tau) d\tau,
	\label{eq:non-local}
\end{align}the non-local generator $\mathcal{K}(t)$ may be given an alternate representation via the memory kernel function $k(t)$ as
	\begin{align}
		\mathcal{K}(t) = k(t)[\mathcal{P}-\mathds{1}],
		\label{eq:memoryker}
	\end{align}
	where the nature of the memory kernel function $k(t)$ determines if the dynamics is non-Markovian or not. The non-local equation for the map gives rise to the solution 
	\begin{align}
	\Phi(t) = g(t) \mathds{1} + (1-g(t)) \mathcal{P},
	\label{eq:projector_map}
	\end{align}with $g(t) = 1 - \int_{0}^{t} d\tau f(\tau)$, where $f(t)$ is associated with the memory kernel $k(t)$ via the Laplace transform $\tilde{f}(t) = \int_{0}^{t}  d\tau e^{-ut} f(\tau)$ such that
	\begin{align}
		\tilde{k}(t) = \frac{u \tilde{f}(t)}{1-\tilde{f}(t)}.
	\end{align}
The dynamical map $\Phi(t)$ representing a quantum process of the above type (\ref{eq:memoryker}), is said to be semi-Markov if $f(t) \ge 0$ and $\int_{0}^{\infty} d\tau f(\tau) \le 1$ \cite{wudarski2016markovian}.
	
	On the other hand,  if the two-time correlation function of the environment spectral density is delta correlated then the Born-Markov approximation holds with the $\mathcal{K}(\tau) = 2 \delta(\tau)\mathcal{L}$, giving rise to semigroup evolution \cite{breuer2002theory} and the dynamical map is given by $\Phi(t) = \exp(t \mathcal{L})$. \bla

	\section{Example 1: Semi-Markov dephasing}

	\subsection*{Semi-Markov dephasing with convolution between \textit{same} waiting time distributions}
As a first example, we consider a dephasing quantum semi-Markov process given by the operator-sum representation $\Phi(t)[\rho] = \sum_j \mathcal{C}_j \rho \mathcal{C}^\dagger _j $, with the Kraus operators
\begin{align}
\mathcal{C}_1 = \sqrt{\frac{1+q(t)}{2}}\mathds{1} \quad ; \quad \mathcal{C}_2 = \sqrt{\frac{1-q(t)}{2}}Z,
\label{eq:semiRTN}
\end{align}
with 
\begin{align}
q(t) = e^{-st/2}\bigg(\cosh\bigg[\frac{\eta st}{2}\bigg]+ \frac{1}{\eta}\sinh\bigg[\frac{\eta  st}{2}\bigg]\bigg),
\label{eq:qoft}
\end{align} where
\begin{align}
\eta = \sqrt{1 - 8 \frac{p}{s^2}}. \label{eq:eta}
\end{align}

The function $q(t)$ is obtained by considering a convolution $\mathsf{f}(t) = f(t) \ast f(t)$ of two exponential (memoryless) waiting time distributions having the form $f(t) = \lambda e^{-\lambda t}$, which is a special instance of the Erlang distribution, with shape parameter $k \equiv 1$. Here $\lambda$ has the physical interpretation of rates determining the speed of the process \cite{vacchini2011markovianity}. Note that a delta correlated memory kernel $k(t)$ with $q(t)$ taking an exponential form ($q(t) = e^{- 2\lambda t}$), leads to a semigroup evolution with constant decay rate $\lambda$.   In the case below which we consider, the two convolved processes are identical, so that the parameters are $s = 2 \lambda$ and $p = \lambda^2$ in the expression for $q(t)$. Interesting point here to note is that a convolution of two \textit{same} exponential waiting time distribution need not result in a CP-divisible evolution. In fact, the example below with convolution of same functions $f_i$ leads to not only CP-indivisible process but also P-indivisible process, which can be straightforwardly verified by the distinguishability measure \cite{breuer2009measure,vacchini2011markovianity}. In fact, for a process involving a single decoherence channel, P-indivisibility and CP-indivisibility are equivalent \cite{vacchini2011markovianity}.

	For $p < \frac{s^2}{8}$, the process is said to be Markovian according to divisibility and distinguishability criteria even though it is non-Markovian classically. It is CP-indivisible when $p > \frac{s^2}{8}$. When $p \rightarrow 0$, we have a QDS limit, that is $q(t) = 1$. 
	
In the time-local master equation 
	\begin{align}
		\frac{d\rho }{dt} = \gamma(t)[\sigma_z \rho \sigma_z - \rho], \label{eq:dephasingmaster}
	\end{align} describing the dephasing process as given above, the decay rates given by \cite{shrikant2018non-Markovian,vacchini2011markovianity}
\begin{align}
	\gamma(t) = -\frac{1}{2}\frac{d}{dt} \big(\ln q(t) \big). 
\end{align}
where $q(t)$ is given in Eq. (\ref{eq:qoft}). For the particular evolution we consider, the  decay rate in the semigroup limit turns out to be $\gamma = 0$, i.e., the identity map, which is trivially a semigroup.   We have plotted the SSS measure for this process in the Figure (\ref{fig:semi-plot}), corresponding to the right divided region. 



\subsection*{Semi-Markov dephasing with convolution between \textit{different} waiting time distributions}
	Earlier we noted that convolution of same waiting time distribution gives rise to CP-indivisible process, whose decay rate consisted of both negative (CP-indivisible) and positive (CP-divisible) regions, which were captured by our measure. Interestingly, convolution between two different waiting time distribution for the earlier dephasing process leads to a master equation of the form $\mathcal{L}[\rho] = \gamma(t)(\sigma_z \rho \sigma_z - \rho)$ with \cite{vacchini2011markovianity}
	\begin{align}
	\gamma(t) = \frac{2p}{s} \frac{1}{1+\eta \coth(\frac{st}{2} \eta)}
	\end{align}
	where $q(t)$ and $\eta$ are given as before in Eqs. (\ref{eq:qoft}) and (\ref{eq:eta}), but only with the change that the function $q(t)$ is now obtained by considering a convolution $\mathsf{f}(t) = f_1(t) \ast f_2(t)$ of two exponential (memoryless) waiting time distributions of the form $f_1(t) = \lambda_1 e^{-\lambda_1 t}$ and $f_2(t) = \lambda_2 e^{-\lambda_2 t}$, where $\lambda_i$ have the physical interpretation of rates determining the speed of the process and the parameters are set to $s= \lambda_1 + \lambda_2$ and $p = \lambda_1 \lambda_2$ \cite{vacchini2011markovianity}. 
	
	Note that this semi-Markov process is CP-divisible with all-time positive decay rate as the condition that the convolution of different waiting time corresponds to the region $ p \ge \frac{s^2}{8}$. However, it is CP-divisible quantum non-Markovian according to our measure corresponding to the left divided region in Figure (\ref{fig:semi-plot}).
	
Now we show that this dephasing semi-Markov process is "quantum" non-Markovian even in the CP-divisible region according to the measure introduced in \cite{shrikant2020temporal}, and we will also quantify the non-Markovianity in the CP-indivisible region. 
	
	\subsection*{Quantifying non-Markovianity of quantum semi-Markov dephasing \label{sec:sec4}} 
	
	Note that the dephasing process in Eq. (\ref{eq:semiRTN}) in the CP-indivisible region, i.e., $p > \frac{s^2}{8}$, corresponds to a singular generator at the master equation level. The decay rate for the process (\ref{eq:semiRTN}) is obtained as
	\begin{align}
	\gamma(t) =\frac{2 p}{s \sqrt{1-\frac{8 p}{s^2}} \coth \left(\frac{1}{2} s t \sqrt{1-\frac{8 p}{s^2}}\right)+s}.
	\label{eq:rate}
	\end{align}
	
\begin{figure}
		\centering
		\includegraphics[width=0.4\textwidth]{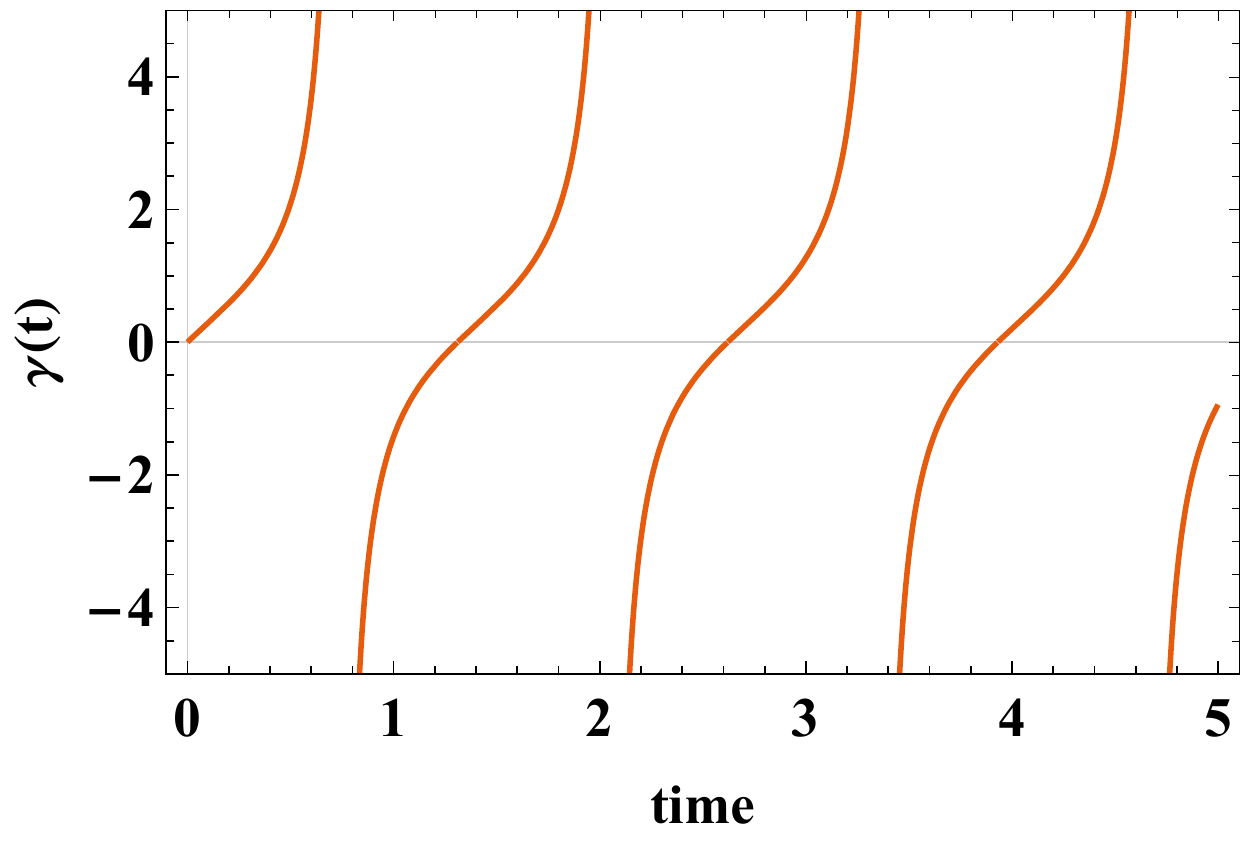}
		\caption{ (Color online) Plot of the decay rate of the semi-Markov process (\ref{eq:semiRTN}), in the CP-indivisible regime.}
		\label{fig:semi-rate} 
	\end{figure}

\begin{figure}
		\centering
		\includegraphics[width=0.45\textwidth]{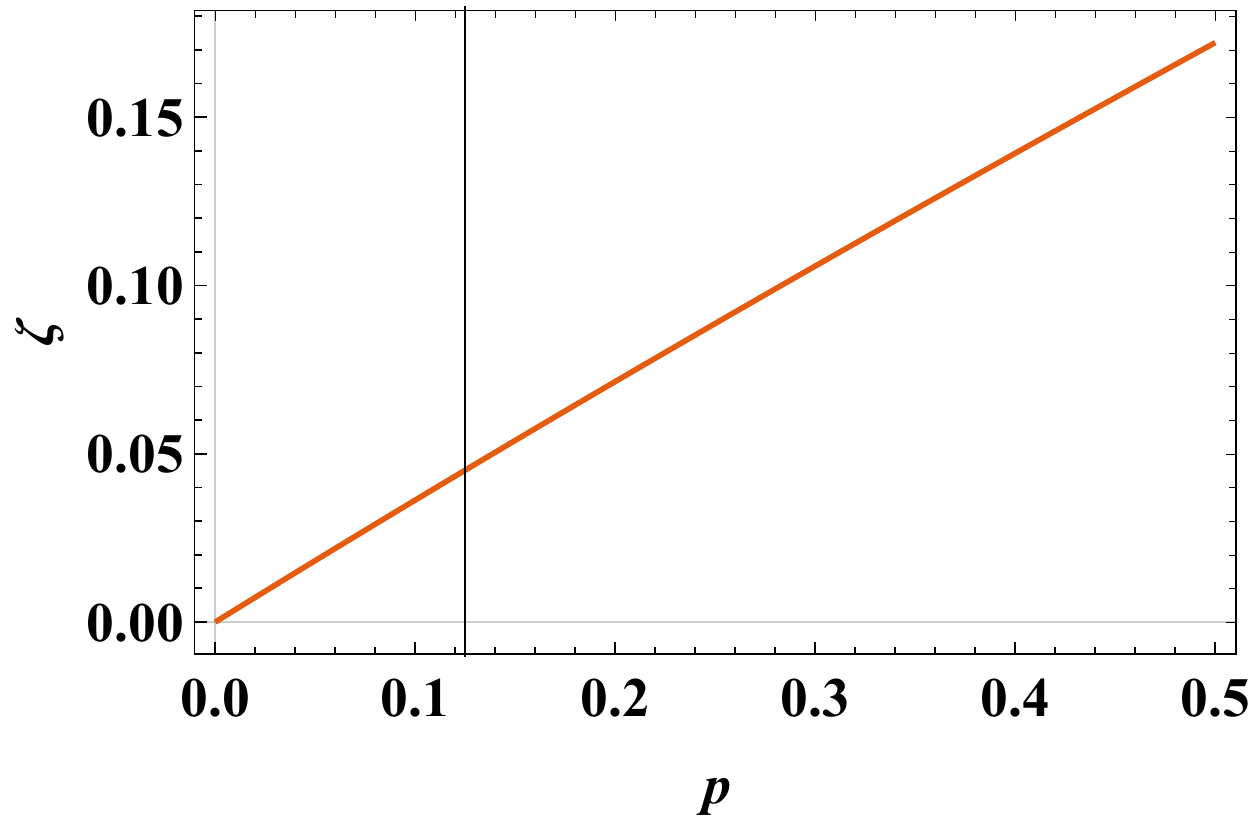}
		\caption{ (Color online) Normalized ($\zeta \equiv \frac{\xi}{1+\xi}$) non-Markovianity measure $\zeta$ (of the SSS measure $\xi$ given in Eqs. (\ref{eq:fqds}) and (\ref{eq:ddimdephase})) with respect to the parameter $p$, which provides a unified description of the CP-divisible region corresponding to \textit{different} waiting time distributions i.e.,  $(\lambda_1 \ne \lambda_2)$, and CP-indivisible region corresponding to \textit{same} waiting time distributions $(\lambda_1 = \lambda_2)$ regimes of the QSM process. Here $s=1$, with a fixed time-interval of $T=1$ and  for the parameter $p$, we consider the range $\{0,\frac{1}{2}\}$.  We can show that the semigroup form corresponds to the point where $p = 0$. That is, when $p$ is this value, then $\gamma = 0$ (a constant, for a given value of $s$).   The figure illustrates the effects of deviation from QDS; when $p > s^2/8$ (for $s=1$), we get the regime where its non-Markovian by the standard measures. The vertical line at $p= \frac{1}{8}$ divides the CP-divisible and CP-indivisible regions of the quantum semi-Markov processes.}
		\label{fig:semi-plot} 
	\end{figure}

	For $p < \frac{s^2}{8}$, the process is CP-divisible. Considering the case of $s=1$, for $p =3 > \frac{1}{8}$ the process is CP-indivisible and the decay rate is plotted in Figure (\ref{fig:semi-rate}). \bla
	 Since the range of $p$ is unbounded after $\frac{s^2}{8}$, the measure Eq. (\ref{eq:fqds}) goes to infinity as $p$ goes to infinity. Therefore, a new measure must be considered that is the normalized version of $\xi$ defined in Eq. (\ref{eq:fqds}) and fits in the range $\{0,1\}$. A convenient renormalized measure would be $\zeta = \frac{\xi}{1+\xi}$. The figure (\ref{fig:semi-plot}) shows the deviation of the semi-Markov process from QDS structure for $s=1$, for the range $p \in [0,\frac{1}{2}]$. This brings out the non-Markovian behavior of semi-Markov processes, even in the CP-divisible regime.
	
	\subsection*{Decoherence mitigation in semi-Markov dephasing processes \label{sec:sec3}}
	\begin{figure}
		\centering
		\includegraphics[width=0.5\textwidth]{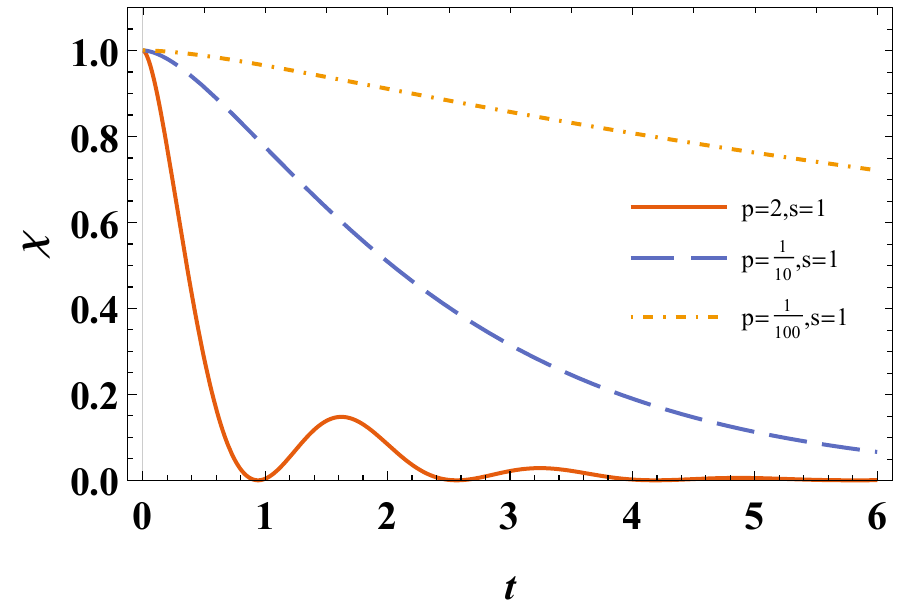}
		\caption{ (Color online) The Holevo information $\chi$ as a function of time for various values of the parameter $p$, which determines the appropriate regions. Here, $s=1$ in all the considered cases. The bold, orange curve corresponds to CP-indivisible region with $p=2$; the blue, dashed curve to semi-Markov with $p=\frac{1}{10}$; and yellow dot-dashed curve to the semigroup evolution with $p=\frac{1}{100}$.}
		\label{fig:semi-holevo} 
	\end{figure}
	Given the quantum states $\rho_j$ occuring with probability $p_j$, the maximum amount of classical information that can be encoded using them is \bla upper bounded by the Holevo $\chi$ quantity $
	\chi := S (\rho ) - \sum_k p_k S(\rho_k)$. The quantity $ S(\rho) = - {\rm Tr}(\rho \log \rho)$, is the von Neumann entropy.

	Keeping in mind that information contained in the quantum states decay under decoherence, one may analyze the effect of semi-Markov processes on the Holevo bound which gives an upper bound on the accessible information. In \cite{fanchini2014non-Markovianity}, accessible  information was used to re-instate the interpretation of information back-flow, first discovered in \cite{breuer2009measure}, with non-Markovianity of the quantum evolution.  As is known, information back-flow is associated with the recurrence of distinguishability of two initially orthogonal states under non-Markovian CPTP maps. Figure (\ref{fig:semi-holevo}) depicts the decay of Holevo information under the semi-Markov processes in the semigroup limit, and in the CP-divisible and CP-indivisible regimes. It is worth noting that, semi-Markov processes in the CP-divisible regime can mitigate the loss of information by slowing down the rate of information leakage into the environment, i.e., slowing down the rate of decrease of distinguishability, while CP-indivisible processes can, for certain intervals, revive the same. In that, we observe that slowing down of the decoherence corresponds to a semi-Markov process with \textit{different} parameters in the convolution product of two waiting time distributions, while the revivals of lost information corresponds to convolution of the waiting time distributions with same parameters. This provides an operational meaning to the non-Markovian behavior of quantum semi-Markov processes. It is important to note that semigroup evolution corresponds to weaker system-environment coupling, hence it is natural that the decay of information is very slow. In contrast to this, one expects that the decay is rapid when the coupling is strong enough. In this light, semi-Markov processes may be thought to be intermediate between the two extreme cases of semigroup and strong non-Markovian behavior, which manifests itself by mitigating the loss of information. One should note that semi-Markovian processes can also be non-Markovian according to the measure due to Breuer-Laine-Piilo (BLP) \cite{breuer2009measure}. In fact, the dynamics considered in Eq.(\ref{eq:semiRTN}) is BLP non-Markovian in the CP-indivisible region. This could be envisaged to have potential applications in quantum information processing.  In order to harness the advantages due to semi-Markov nature of the noise, one may consider a possibility of simulating the noisy evolution of the qubit. See for example, \cite{cialdi2017all} where an all-optical simulation of random telegraph noise is given. 

\section{Example 2: Semi-Markov non-unital channel}
Here we consider a non-unital dynamics of a qubit as follows. Following Eq.(\ref{eq:projector_map}), the evolution map may be given by 
\begin{align}
\Phi(t) =  [1-\int_{0}^{t} d\tau f(\tau)] \mathds{1} + [\int_{0}^{t} d\tau f(\tau)] \mathcal{P}
\label{eq:damp-projector0}
\end{align}
where $\mathcal{P}$ is the CPTP projector defined by 
\begin{align}
\mathcal{P}[\rho] = \vert 0 \rangle \langle 0 \vert \text{Tr}(\rho)
\label{eq:damp-projector}
\end{align} 
The process Eq. (\ref{eq:damp-projector0}) can be interpreted as a quantum particle on a lattice with damped hopping, with waiting time distribution given by $f(t)$ mentioned before and the survival probability of the particle given by $g(t) = 1 - \int_{0}^{t} d\tau f(\tau)$. Consider a waiting time distribution of the form
\begin{align}
f(t) = \lambda \tanh(\lambda t) \sech(\lambda t)
 \label{eq:waiting_time_damping2}
\end{align} 
where $\lambda$ is a eal number. Note that $f(t) \ge 0$ and $\int_{0}^{\infty} f(t) \le 1$ for all $t$ and for $\lambda > 0$, hence it is a semi-Markov process. The master equation is obtained straightforwardly:
\begin{align}
\mathcal{L}(t)= \lambda \tanh(\lambda t)[\mathcal{P}- \mathds{1}],
\end{align} where $\mathcal{P}$ is given by Eq.(\ref{eq:damp-projector}). 
Here $\lambda$ parametrizes a family with the limit $\lambda {\rightarrow} 0$ corresponding to the identity map,
Clearly, the dynamical map $\Phi$ for this process deviates from semigroup evolution as it is not constant; furthermore, as the Lindblad rate $\gamma(t) {:=} \lambda \tanh(\lambda t) > 0$ for all $t > 0$ and for $\lambda > 0$, the processes CP-divisible. We will now show that it is non-Markovian process according to our measure. 
In this family, the QDS limit is the identity map, obtained by setting $\lambda \rightarrow 0$. which is of course the trivial semigroup. In this example of a semi-Markov process, there is no CP-indivisible region, unlike the previous one. (CP-indivisible regions can be introduced by letting $\lambda$ take imaginary values, which can be implemented by a slightly different parametrization scheme).

From Eq.(\ref{eq:fqds}), the SSS measure for this process turns out to be simply
\begin{align}
\xi = \frac{1}{T} \int_{0}^{t} \vert \gamma(t) - \gamma \vert dt \nonumber
\end{align} where $\gamma(t)=\lambda \tanh(\lambda t)$ for the CP-divisible non-Markovian regime and $\gamma = 0$ is the semigroup rate for this process. Therefore, for a simple case of $T=1$, the measure becomes
\begin{align}
\xi = \int_{0}^{1} \lambda \tanh(\lambda t) dt = \ln \cosh(\lambda). \nonumber
\end{align}
As one increases the parameter $\lambda$, the process becomes more and more non-Markovian according to SSS measure.

	\section{Conclusions \label{sec:sec5}}
	Here, we have studied an interesting stochastic process called the semi-Markov process. This belongs to a category of processes that exhibit memory effects in the CP-divisible regime. The principal tool used to quantify this memory, in the quantum regime, is the recently introduced  SSS measure \cite{shrikant2020temporal},  that quantifies the non-Markovianity  as a function of the magnitute of deviation from semigroup evolution. Further, the behavior of  the Holevo bound under the influence of semi-Markov processes brings out the effect of decoherence mitigation, that is,  the loss of information leakage into the environment is slowed down. This provides an operational meaning to the non-Markovian  nature of semi-Markov processes. 
	
As was noted earlier \cite{shrikant2020temporal}, PL and OU noises are non-Markovian according to the SSS criterion,  and yet CP-divisible. In that light, it would be an interesting future direction to explore the relation between semi-Markov dynamics and such CP-divisible but SSS non-Markovian processes such as PL and OU noises, i.e., to study whether it is generally the case that dynamics that is CP-divisible but non-Markovian in the SSS sense, falls in the category of quantum semi-Markov processes. 
	
	
\section*{Acknowledgments}
SU acknowledges the support from Admar Mutt Education Foundation, Udupi for scholarship. RS acknowledges the support of Department of Science and Technology (DST), India, Grant No.: MTR/2019/001516.
	
\bibliography{semi-Markov}

\begin{thebibliography}{10}

\bibitem{banerjee2018open}
Subhashish Banerjee.
\newblock {\em Open quantum systems}.
\newblock Springer Nature Singapore Pte Ltd., 2018.

\bibitem{breuer2002theory}
Heinz-Peter Breuer and Francesco Petruccione.
\newblock {\em The theory of open quantum systems}.
\newblock Oxford University Press, 2002.

\bibitem{grabert1988quantum}
Hermann Grabert, Peter Schramm, and Gert-Ludwig Ingold.
\newblock Quantum brownian motion: The functional integral approach.
\newblock {\em Physics Reports}, 168(3):115 -- 207, 1988.

\bibitem{sbqbm}
Subhashish Banerjee and R.~Ghosh.
\newblock General quantum brownian motion with initially correlated and
  nonlinearly coupled environment.
\newblock {\em Phys. Rev. E}, 67:056120, May 2003.

\bibitem{banerjee2000quantum}
Subhashish Banerjee and R~Ghosh.
\newblock Quantum theory of a stern-gerlach system in contact with a linearly
  dissipative environment.
\newblock {\em Phys. Rev. A}, 62(4):042105, 2000.

\bibitem{RHP14}
Angel Rivas, Susana~F Huelga, and Martin~B Plenio.
\newblock Quantum non-markovianity: characterization, quantification and
  detection.
\newblock {\em Rep. Prog. Phys}, 77(9):094001, 2014.

\bibitem{breuer2016colloquium}
Heinz-Peter Breuer, Elsi-Mari Laine, Jyrki Piilo, and Bassano Vacchini.
\newblock Colloquium: Non-markovian dynamics in open quantum systems.
\newblock {\em Rev. Mod. Phys}, 88(2):021002, 2016.

\bibitem{vega2017dynamics}
In\'es de~Vega and Daniel Alonso.
\newblock Dynamics of non-markovian open quantum systems.
\newblock {\em Rev. Mod. Phys.}, 89:015001, Jan 2017.

\bibitem{li2018concepts}
Li~Li, Michael~J.W. Hall, and Howard~M. Wiseman.
\newblock Concepts of quantum non-markovianity: A hierarchy.
\newblock {\em Physics Reports}, 759:1 -- 51, 2018.

\bibitem{shrikant2020temporal}
Shrikant Utagi, R~Srikanth, and Subhashish Banerjee.
\newblock Temporal self-similarity of quantum dynamical maps as a concept of
  memorylessness.
\newblock {\em Scientific Reports}, 10(1):1--10, 2020.

\bibitem{shrikant2018non-Markovian}
U.~Shrikant, R.~Srikanth, and Subhashish Banerjee.
\newblock Non-markovian dephasing and depolarizing channels.
\newblock {\em Phys. Rev. A}, 98:032328, Sep 2018.

\bibitem{vasile2011continuous}
Ruggero Vasile, Stefano Olivares, Matteo G.~A. Paris, and Sabrina Maniscalco.
\newblock Continuous-variable quantum key distribution in non-markovian
  channels.
\newblock {\em Phys. Rev. A}, 83:042321, Apr 2011.

\bibitem{thapliyal2017quantum}
Kishore Thapliyal, Anirban Pathak, and Subhashish Banerjee.
\newblock Quantum cryptography over non-markovian channels.
\newblock {\em Quantum Information Processing}, 16(5):115, 2017.

\bibitem{shrikant2020pingpong}
Shrikant Utagi, R~Srikanth, and Subhashish Banerjee.
\newblock Ping-pong quantum key distribution with trusted noise: non-markovian
  advantage.
\newblock {\em Quantum Information Processing}, 19(366), 2020.

\bibitem{kumar2017nonmarkovian}
N.~Pradeep Kumar, Subhashish Banerjee, R.~Srikanth, Vinayak Jagadish, and
  Francesco Petruccione.
\newblock Non-markovian evolution: a quantum walk perspective.
\newblock {\em Open systems \& Information Dynam}, 25(03):1850014, 2018.

\bibitem{naikoo2020non}
Javid Naikoo, Subhashish Banerjee, and CM~Chandrashekar.
\newblock Non-markovian channel from the reduced dynamics of a coin in a
  quantum walk.
\newblock {\em Physical Review A}, 102(6):062209, 2020.

\bibitem{thomas2018thermodynamics}
George Thomas, Nana Siddharth, Subhashish Banerjee, and Sibasish Ghosh.
\newblock Thermodynamics of non-markovian reservoirs and heat engines.
\newblock {\em Phys. Rev. E}, 97:062108, Jun 2018.

\bibitem{laine2014nonlocal}
Elsi-Mari Laine, Heinz-Peter Breuer, and Jyrki Piilo.
\newblock Nonlocal memory effects allow perfect teleportation with mixed
  states.
\newblock {\em Scientific reports}, 4(1):1--5, 2014.

\bibitem{bhattacharya2018evolution}
Samyadeb Bhattacharya, Subhashish Banerjee, and Arun~Kumar Pati.
\newblock Evolution of coherence and non-classicality under global
  environmental interaction.
\newblock {\em Quantum Information Processing}, 17(9):236, 2018.

\bibitem{banerjee2010dynamics}
Subhashish Banerjee, V~Ravishankar, and R~Srikanth.
\newblock Dynamics of entanglement in two-qubit open system interacting with a
  squeezed thermal bath via dissipative interaction.
\newblock {\em Annals of Physics}, 325(4):816--834, 2010.

\bibitem{paulson2020hierarchy}
KG~Paulson, Ekta Panwar, Subhashish Banerjee, and R~Srikanth.
\newblock Hierarchy of quantum correlations under non-markovian dynamics.
\newblock {\em arXiv preprint arXiv:2004.11208}, 2020.

\bibitem{RHP10}
{\'A}ngel Rivas, Susana~F Huelga, and Martin~B Plenio.
\newblock Entanglement and non-markovianity of quantum evolutions.
\newblock {\em Phys. Rev. Lett}, 105(5):050403, 2010.

\bibitem{lindblad1976generators}
Goran Lindblad.
\newblock On the generators of quantum dynamical semigroups.
\newblock {\em Communications in Mathematical Physics}, 48(2):119--130, 1976.

\bibitem{gorini1976completely}
Vittorio Gorini, Andrzej Kossakowski, and Ennackal Chandy~George Sudarshan.
\newblock Completely positive dynamical semigroups of n-level systems.
\newblock {\em Journal of Mathematical Physics}, 17(5):821--825, 1976.

\bibitem{vacchini2014general}
Bassano Vacchini.
\newblock General structure of quantum collisional models.
\newblock {\em International Journal of Quantum Information}, 12(02):1461011,
  2014.

\bibitem{ciccarello2013collision}
F.~Ciccarello, G.~M. Palma, and V.~Giovannetti.
\newblock Collision-model-based approach to non-markovian quantum dynamics.
\newblock {\em Phys. Rev. A}, 87:040103, Apr 2013.

\bibitem{pollock2018operational}
Felix~A. Pollock, C\'esar Rodr\'{\i}guez-Rosario, Thomas Frauenheim, Mauro
  Paternostro, and Kavan Modi.
\newblock Operational markov condition for quantum processes.
\newblock {\em Phys. Rev. Lett.}, 120:040405, Jan 2018.

\bibitem{shabani2005completely}
A.~Shabani and D.~A. Lidar.
\newblock Completely positive post-markovian master equation via a measurement
  approach.
\newblock {\em Phys. Rev. A}, 71:020101, Feb 2005.

\bibitem{breuer2009measure}
Heinz-Peter Breuer, Elsi-Mari Laine, and Jyrki Piilo.
\newblock Measure for the degree of non-markovian behavior of quantum processes
  in open systems.
\newblock {\em Phys. Rev. Lett}, 103(21):210401, 2009.

\bibitem{budini2018quantum}
Adri\'an~A. Budini.
\newblock Quantum non-markovian processes break conditional past-future
  independence.
\newblock {\em Phys. Rev. Lett.}, 121:240401, Dec 2018.

\bibitem{budini2019conditional}
Adri\'an~A. Budini.
\newblock Conditional past-future correlation induced by non-markovian
  dephasing reservoirs.
\newblock {\em Phys. Rev. A}, 99:052125, May 2019.

\bibitem{chruscinski2014degree}
Dariusz Chru{\'s}ci{\'n}ski and Sabrina Maniscalco.
\newblock Degree of non-markovianity of quantum evolution.
\newblock {\em Physical review letters}, 112(12):120404, 2014.

\bibitem{davalos2019divisibility}
David Davalos, Mario Ziman, and Carlos Pineda.
\newblock Divisibility of qubit channels and dynamical maps.
\newblock {\em Quantum}, 3:144, 2019.

\bibitem{breuer2009structure}
Heinz-Peter Breuer and Bassano Vacchini.
\newblock Structure of completely positive quantum master equations with memory
  kernel.
\newblock {\em Physical Review E}, 79(4):041147, 2009.

\bibitem{megier2021memory}
Nina Megier, Manuel Ponzi, Andrea Smirne, and Bassano Vacchini.
\newblock Memory effects in quantum dynamics modelled by quantum renewal
  processes.
\newblock {\em Entropy}, 23:905, 2021.

\bibitem{yu2010entanglement}
Ting Yu and J.H. Eberly.
\newblock Entanglement evolution in a non-markovian environment.
\newblock {\em Optics Communications}, 283(5):676 -- 680, 2010.

\bibitem{hall2014canonical}
Michael J.~W. Hall, James~D. Cresser, Li~Li, and Erika Andersson.
\newblock Canonical form of master equations and characterization of
  non-markovianity.
\newblock {\em Phys. Rev. A}, 89:042120, Apr 2014.

\bibitem{chrusinski2018divisibility}
Dariusz Chru\ifmmode \acute{s}\else \'{s}\fi{}ci\ifmmode~\acute{n}\else
  \'{n}\fi{}ski, \'Angel Rivas, and Erling St\o{}rmer.
\newblock Divisibility and information flow notions of quantum markovianity for
  noninvertible dynamical maps.
\newblock {\em Phys. Rev. Lett.}, 121:080407, Aug 2018.

\bibitem{vacchini2011markovianity}
Bassano Vacchini, Andrea Smirne, Elsi-Mari Laine, Jyrki Piilo, and Heinz-Peter
  Breuer.
\newblock Markovianity and non-markovianity in quantum and classical systems.
\newblock {\em New Journal of Physics}, 13(9):093004, 2011.

\bibitem{megier2020evolution}
Nina Megier, Andrea Smirne, and Bassano Vacchini.
\newblock Evolution equations for quantum semi-markov dynamics.
\newblock {\em Entropy}, 22(7):796, 2020.

\bibitem{wudarski2016markovian}
Filip~A Wudarski and Dariusz Chru{\'s}ci{\'n}ski.
\newblock Markovian semigroup from non-markovian evolutions.
\newblock {\em Physical Review A}, 93(4):042120, 2016.

\bibitem{fanchini2014non-Markovianity}
Felipe~F Fanchini, Goktug Karpat, Baris {\c{C}}akmak, LK~Castelano, GH~Aguilar,
  O~Jim{\'e}nez Far{\'\i}as, SP~Walborn, PH~Souto Ribeiro, and MC~De~Oliveira.
\newblock Non-markovianity through accessible information.
\newblock {\em Physical Review Letters}, 112(21):210402, 2014.

\bibitem{cialdi2017all}
Simone Cialdi, Matteo~AC Rossi, Claudia Benedetti, Bassano Vacchini, Dario
  Tamascelli, Stefano Olivares, and Matteo~GA Paris.
\newblock All-optical quantum simulator of qubit noisy channels.
\newblock {\em Applied Physics Letters}, 110(8):081107, 2017.

\end{thebibliography}

\end{document}